\documentclass[%
 reprint,
 superscriptaddress,
 amsmath,amssymb,
 aps,
 pra,
floatfix,
]{revtex4-1}

\usepackage{graphicx}% Include figure files
\usepackage{dcolumn}% Align table columns on decimal point
\usepackage{bm}% bold math
\usepackage{siunitx}
\usepackage[usenames, dvipsnames]{xcolor}
\usepackage{hyperref}% add hypertext capabilities
\begin{document}
%%%%%%%%%%%%%%%%%%%%%%%%%%%%%%%%%%%%%%%%%%%%%%%%%%%%%%%%%%%%%
\title{Autotuning of double dot devices {\it in situ} with machine learning}

\author{Justyna P. Zwolak}
\email{jpzwolak@nist.gov}
\affiliation{National Institute of Standards and Technology, Gaithersburg, Maryland 20899, USA}

\author{Thomas McJunkin}
\email{tmcjunkin@wisc.edu}
\affiliation{Department of Physics, University of Wisconsin-Madison, Madison, Wisconsin 53706, USA}

\author{Sandesh S. Kalantre}
\affiliation{Joint Quantum Institute, University of Maryland, College Park, Maryland 20742, USA}
\affiliation{Joint Center for Quantum Information and Computer Science, 
University of Maryland, College Park, Maryland 20742, USA}

\author{J. P. Dodson}
\affiliation{Department of Physics, University of Wisconsin-Madison, Madison, Wisconsin 53706, USA}

\author{E. R. MacQuarrie}
\affiliation{Department of Physics, University of Wisconsin-Madison, Madison, Wisconsin 53706, USA}

\author{D. E. Savage}
\affiliation{Department of Materials Science and Engineering, University of Wisconsin-Madison, Madison, Wisconsin 53706, USA}

\author{M. G. Lagally}
\affiliation{Department of Materials Science and Engineering, University of Wisconsin-Madison, Madison, Wisconsin 53706, USA}

\author{S. N. Coppersmith}
\affiliation{Department of Physics, University of Wisconsin-Madison, Madison, Wisconsin 53706, USA}
\affiliation{School of Physics, The University of New South Wales, Sydney, New South Wales, Australia}

\author{Mark A. Eriksson}
\affiliation{Department of Physics, University of Wisconsin-Madison, Madison, Wisconsin 53706, USA}

\author{Jacob M. Taylor}
\affiliation{National Institute of Standards and Technology, Gaithersburg, Maryland 20899, USA}
\affiliation{Joint Quantum Institute, University of Maryland, College Park, Maryland 20742, USA}
\affiliation{Joint Center for Quantum Information and Computer Science,
University of Maryland, College Park, Maryland 20742, USA}

\date{\today}
%%%%%%%%%%%%%%%%%%%%%%%%%%%%%%%%%%%%%%%%%%%%%%%%%%%%%%%%%%%%%
\begin{abstract}
The current practice of manually tuning quantum dots (QDs) for qubit operation is a relatively time-consuming procedure that is inherently impractical for scaling up and applications. In this work, we report on the {\it in situ} implementation of a recently proposed autotuning protocol that combines machine learning (ML) with an optimization routine to navigate the parameter space. In particular, we show that a ML algorithm trained using exclusively simulated data to quantitatively classify the state of a double-QD device can be used to replace human heuristics in the tuning of gate voltages in real devices. We demonstrate active feedback of a functional double-dot device operated at millikelvin temperatures and discuss success rates as a function of the initial conditions and the device performance. Modifications to the training network, fitness function, and optimizer are discussed as a path toward further improvement in the success rate when starting both near and far detuned from the target double-dot range.
\end{abstract}
\maketitle

%%%%%%%%%%%%%%%%%%%%%%%%%%%%%%%%%%%%%%%%%%%%%%%%%%%%%%%%%%%%%
%---INTORDUCTION---%
%%%%%%%%%%%%%%%%%%%%%%%%%%%%%%%%%%%%%%%%%%%%%%%%%%%%%%%%%%%%%
\section{Introduction}\label{sec:intro}
%%%%%%%%%%%%%%%%%%%%%%%%%%%%%%%%%%%%%%%%%%%%%%%%%%%%%%%%%%%%%
Arrays of quantum dots (QDs) are one of many candidate systems to realize qubits---the fundamental building blocks of quantum computers---and to provide a platform for quantum computing~\cite{Loss98-QCD,Hanson07-SQD,Zwanenburg13-SQE}. Due to the ease of control of the relevant parameters~\cite{Petta05-CQM,Koppens06-COS,Medford13-QEQ,Kim15-RGQ}, fast measurement of the spin and charge states~\cite{Barthel09-SSM}, long decoherence times~\cite{Veldhorst14-AQD, Kawakami14-LLQ,Yoneda18-QDC}, and recent demonstration of two-qubit gates and algorithms~\cite{Veldhorst15-LGS,Zajac18-RDC,Watson18-PQP}, QDs are gaining popularity as candidate building blocks for solid-state quantum devices. In semiconductor quantum computing, devices now have tens of individual gate voltages that must be carefully set to isolate the system to the single electron regime and to realize good qubit performance. At the same time, even tuning a double QD constitutes a nontrivial task, with each dot being controlled by at least three metallic gates, each of which influences the number of electrons in the dot, the tunnel coupling to the adjacent lead, and the interdot tunnel coupling. The background potential energy, which is disordered by defects and variations in the local composition of the heterostructure, further impedes this process. In order to reach a stable few-electron configuration, current experiments set the input voltages heuristically. However, such an approach does not scale well with growing array sizes, is prone to random errors, and may result in only an acceptable rather than an optimal state. Moreover, with an increasing number of QD qubits, the relevant parameter space grows exponentially, making heuristic control even more challenging.  

Given the recent progress in the physical construction of larger arrays of quantum dots in both one and two dimensions~\cite{Zajac16-SGA,Mukhopadhyay18-2DD}, it is imperative to have a reliable automated protocol to find a stable desirable electron configuration in the dot array, i.e., to automate finding a set of voltages that yield the desired confinement regions (dots) at the intended positions and with the correct number of electrons and couplings, and to do it efficiently. There have been a number of recent proposals on how to achieve these tasks, including computer-supported algorithmic gate voltage control and pattern matching for tuning~\cite{Baart2016-ATD,Botzem2018-TSD,vanDiepen18-ATC,Teske19-MFT,Mills19-CAT} and a machine-learning-guided protocol aimed at reducing the total number of measurements~\cite{Lennon19-EMM}. However, while these tuning approaches to a lesser or greater extent eliminate the need for human intervention, they are tailored to a particular device's design and need to be adjusted if used on a different one. Moreover, most of these approaches focus on fine tuning to the single-electron regime, assuming some level of knowledge about the parameter ranges that lead to a well-controlled qubit system. 

\begin{figure*}[t]
\includegraphics[width=1.0\linewidth]{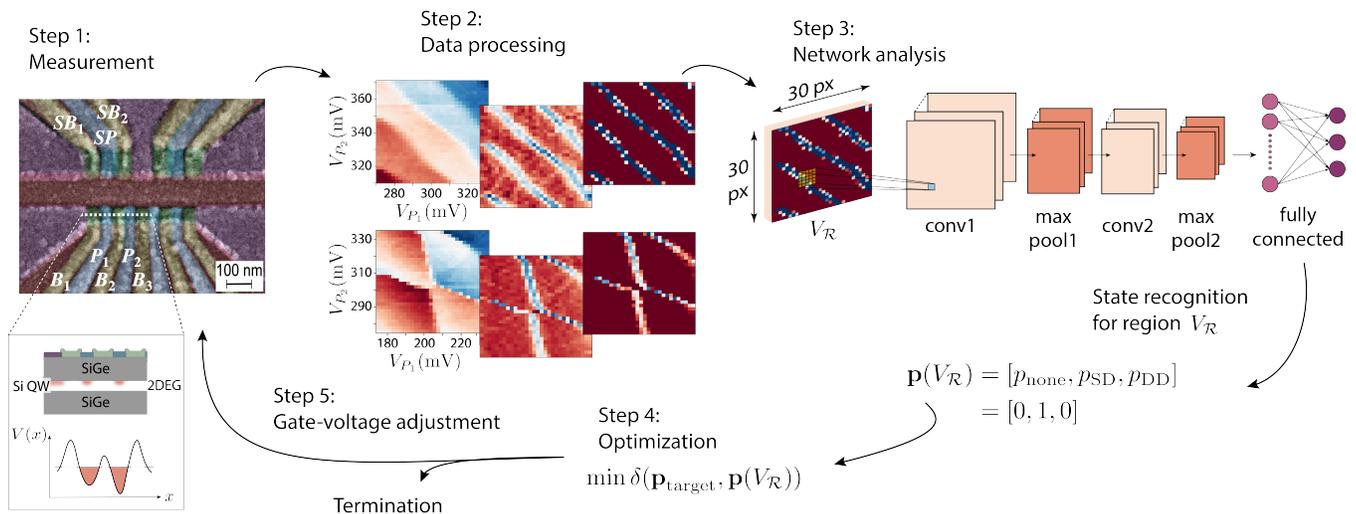}
\caption{Visualization of the autotuning loop. In Step 1, we show a false-color scanning electron micrograph of a Si/Si$_{x}$Ge$_{1-x}$ quadruple-dot device identical to the one measured. The double dot used in the experiment is highlighted by the inset, which shows a cross section through the device along the dashed white line and a schematic of the electric potential of a tuned double dot. ${B_i}$ ($i=1,2,3$) and $P_j$ ($j=1,2$) are the barrier and plunger gates, respectively, used to form dots, while $SB_1$, $SB_2$, and $SP$ are gates (two barriers and a plunger, respectively) used to control the sensing dot. In Step 2, to assure compatibility with the CNN, the raw data are processed and (if necessary) downsized to $(30\times30)$ pixel size. The processed image $V_\mathcal{R}$ is analyzed by the CNN (Step 3), resulting in a probability vector $\text{\bf p}(V_\mathcal{R})$ quantifying the current state of the device. In the optimization phase (Step 4), the algorithm decides whether the state is sufficiently close to the desired one (termination) or whether additional tuning steps are necessary. If the latter, the optimizer returns the position of the consecutive scan (Step 5).}
\label{fig:tuning}
\end{figure*}

Typically, the process of tuning QD devices into qubits involves identifying the global state of the device (e.g., single dot or double dot) from a series of measurements, followed by an adjustment of parameters (gate voltages) based on the observed outcomes. The classification of outcomes can be determined by a trained researcher, identifying the location of QDs based on the relative action of gates and the assembly of multiple QDs based on the relative capacitive shifts. In recent years, machine-learning (ML) algorithms, and specifically convolutional neural networks (CNNs), have emerged as a ``go to'' technique for automated image classification, giving reliable output when trained on a representative and comprehensive data set~\cite{Krizhevsky12-CDN}. Recently, Kalantre {\it et al.} have proposed a new paradigm for fully automated experimental device control---{\it QFlow}---that combines CNNs with optimization techniques to establish a closed-loop control system~\cite{Kalantre17-MLD}. Here, we report on the performance of this autotuning protocol when implemented {\it in situ} on an active quantum-dot device to tune from a single-dot to a double-dot regime. We also discuss further modifications to this protocol to improve overall performance.

The paper is organized as follows: In Sec.~\ref{sec:setup}, we describe the experimental setup. The characteristics of the quantum-dot chip used in the experiment are described in Sect.~\ref{subsec:device}. Overviews of the ML and optimization techniques implemented in the autotuning protocol are presented in Secs.~\ref{subsec:ml} and~\ref{subsec:autotuning}, respectively. The {\it in situ} performance of the autotuner is discussed in Sec.~\ref{sec:tuning-insitu} and the ``off-line'' analysis in Sec.~\ref{sec:tuning-offline}. We conclude with a discussion of the potential modifications to further improve the proposed autotuning technique in Sec.~\ref{sec:summary}.

%%%%%%%%%%%%%%%%%%%%%%%%%%%%%%%%%%%%%%%%%%%%%%%%%%%%%%%%%%%%%
\section{Experimental setup}\label{sec:setup}
%%%%%%%%%%%%%%%%%%%%%%%%%%%%%%%%%%%%%%%%%%%%%%%%%%%%%%%%%%%%%
We define ``autotuning'' as a process of finding a range of gate voltages where the device is in a particular ``global configuration'' (i.e., a no-dot, single-dot, or double-dot regime). The main steps of the experimental implementation of the autotuner are presented in Fig.~\ref{fig:tuning}, with each step discussed in detail in the following sections.

\noindent
{\it Step 0: Preparation.} Before the ML systems are engaged, the device is cooled down and the gates are manually checked for response and pinch-off voltages. Furthermore, the charge sensor and the barrier gates are also tuned using traditional techniques.  \\
\noindent
{\it Step 1: Measurement.} A two-dimensional (2D) measurement of the charge-sensor response over a fixed range of gate voltages. The position for the initial measurement (given as a center and a size of the scan in millivolts) is provided by a user.\\
\noindent
{\it Step 2: Data processing.} Resizing of the measured 2D scan $V_\mathcal{R}$ and filtering of the noise (if necessary) to assure compatibility with the neural network. \\
\noindent
{\it Step 3: Network analysis.}	Analysis of the processed data. The CNN identifies the state of the device for $V_\mathcal{R}$ and returns a probability vector $\text{\bf p}(V_\mathcal{R})$ [see Eq.~(\ref{eq:prob_vec})]. \\
\noindent
{\it Step 4: Optimization.}	An optimization of the fitness function $\delta(\text{\bf p}_{\text{target}},\text{\bf p}(V_\mathcal{R}))$, given in Eq.~(\ref{eq:fit-function}), resulting either in a position of the consecutive 2D scan or decision to terminate the autotuning. \\
\noindent
{\it Step 5: Gate-voltage adjustment.}	 An adjustment of the gate voltages as suggested by the optimizer. The position of the consecutive scan is given as a center of the scan  (in millivolts). \\
\noindent

The preparation step results in a range of acceptable voltages for gates, which allows ``sandboxing'' by limiting the two plunger voltages controlled by the autotuning protocol within these ranges to prevent device damage, as well as in establishment of the appropriate voltage level at which the barrier gates are fixed throughout the test runs (precalibration). The charge-sensing dot is also tuned manually at this stage. The sandbox also helps define the size of the regions used for state recognition. Proper scaling of the measurement scans is crucial for meaningful network analysis: scans that are too small may not contain enough features necessary for state classification while scans that are too large may result in probability vectors that are not useful in the optimization phase.  

Steps 1--5 are repeated until the desired global state is reached. In other words, we formulate the autotuning as an optimization problem over the state of the device in the space of gate voltages, where the function to be optimized is a fitness function $\delta$ between probability vectors of the current and the desired measurement outcomes. The autotuning is considered successful if the optimizer converges to a voltage range that gives the expected dot configuration. 

%%%%%%%%%%%%%%%%%%%%%%%%%%%%%%%%%%%%%%%%%%%%%%%%%%%%%%%%%%%%%
\subsection{Device layout and characteristics}\label{subsec:device}
%%%%%%%%%%%%%%%%%%%%%%%%%%%%%%%%%%%%%%%%%%%%%
QDs are defined by electrostatically confining electrons using voltages on metallic gates applied above a 2D electron gas (2DEG) present at the interface of a semiconductor heterostructure. Realization of good qubit performance is achieved via precise electrostatic confinement, band-gap engineering, and dynamically adjusted voltages on nearby electrical gates. A false-color scanning electron micrograph of a Si/Si$_{x}$Ge$_{1-x}$ quadruple-dot device identical to the one measured is shown in Fig.~\ref{fig:tuning}, Step 1. The device is an overlapping, accumulation-style design~\cite{Zajac15-RGA} consisting of three layers of aluminum surface gates, electrically isolated from the heterostructure surface by deposited aluminum oxide. The layers are isolated from each other by the self-oxidation of the aluminum. The inset in Fig.~\ref{fig:tuning} features a schematic cross section of the device, showing where QDs are expected to form and a modeled potential profile along a one-dimensional (1D) channel formed in the 2DEG. The 2DEG, with an electron mobility of 40\,000 \si{cm^{2}\,\volt^{-1}\,\second^{-1}} at {$4.0\times10^{11}\,$}\si{cm^{-2}}, as measured in a Hall bar, is formed approximately 33\,\si{\nano\meter} below the surface at the upper interface of the silicon quantum well. The application of appropriate voltages to the gates defines the QDs by selectively accumulating and depleting regions within the 2DEG. In particular, depletion ``screening'' gates (shown in red in Fig.~\ref{fig:tuning}) are used to define a 1D transport channel in the 2DEG; reservoir gates (shown in purple in Fig.~\ref{fig:tuning}) accumulate electrons into leads with stable chemical potential; plunger gates (shown in blue and labeled $P_j$, $j = 1, 2$, in Fig.~\ref{fig:tuning}) accumulate electrons into quantum dots and shift the chemical potential in the dots relative to the chemical potential of the leads; and, finally, barrier gates (shown in green and labeled $B_i$, $i = 1, 2, 3$, in Fig.~\ref{fig:tuning}) separate the defined quantum dots and control the tunnel rates between dots and to the leads. In other words, the choice of gate voltages determines the number of dots, their position, their coupling, and the number of electrons present in each dot. Across the central screening gate, opposing the main channel of four linear dots, larger quantum dots are formed to act as sensitive charge sensors capable of detecting single-electron transitions of the main channel quantum dots. The measurements are taken in a dilution refrigerator with a base temperature $< 50\,$\si{\milli\kelvin} and in the absence of an applied magnetic field.

\begin{figure*}[tp]
\includegraphics[width=1.0\linewidth]{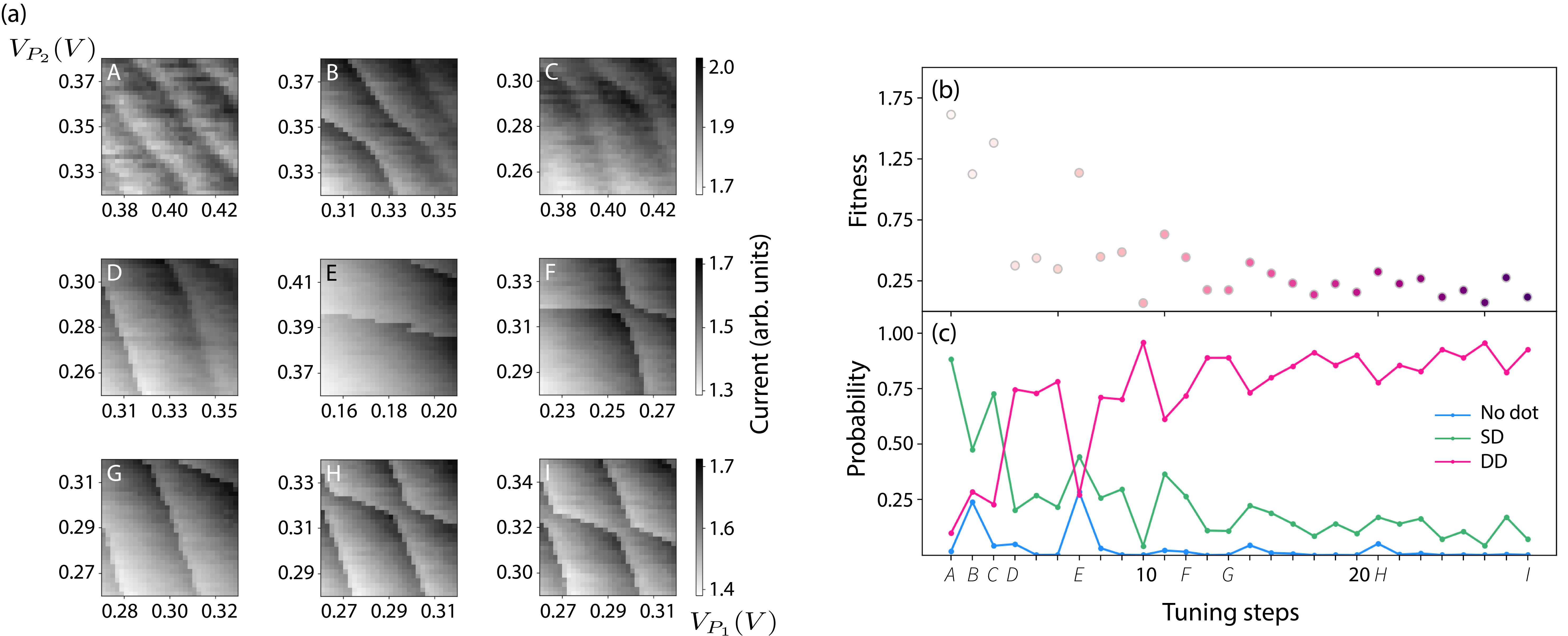}
\caption{A sample run of the autotuning protocol. (a) The measured raw scans in the space of plunger gates $(V_{P_1},V_{P_2})$ show data available to the autotuning protocol at a given time. (b) The change of the fitness value as a function of time. (c) The change in probability of each state over time as returned by the CNN. For an overview of the tuning path in the space of plunger gates on a larger scan measured once the autotuning tests are completed, see Fig.~\ref{fig:tuning_path}.}
\label{fig:sample_run}
\end{figure*}

%%%%%%%%%%%%%%%%%%%%%%%%%%%%%%%%%%%%%%%%%%%%%%%%%%%%%%%%%%%%%
\subsection{Quantitative classification}\label{subsec:ml}
%%%%%%%%%%%%%%%%%%%%%%%%%%%%%%%%%%%%%%%%%%%%%
To automate the tuning process and eliminate the need for human intervention, we incorporate ML techniques into the software controlling the experimental apparatus. In particular, we use a pretrained CNN to determine the current global state of the device. To prepare the CNN, we rely on a data set of 1001 quantum-dot devices generated using a modified Thomas-Fermi approximation to model a set of reference semiconductor systems comprising of a quasi-1D nanowire with a series of depletion gates the voltages of which determine the number of dots, the charges on each of those dots, and the conductance through the wire~\cite{qf-data,Zwolak18-QLD}. The data set is constructed to be agnostic about the details of a particular geometry and material platform used for fabricating dots. To reflect the minimum qualitative features across a wide range of devices, a number of parameters are varied between simulations, such as the device geometry, gate positions, lever arm, and screening length, to name a few. The idea behind varying the device parameters when generating training data set is to enable the use of the same pretrained network on different experimental devices.  

The synthetic data set contains full-size simulated 2D measurements of the charge-sensor readout and the state labels at each point as functions of plunger gate voltages $(V_{P_1},V_{P_2})$ (at a pixel level). For training purposes, we generate an assembly of 10\,010 random charge-sensor measurement realizations (ten samples per full-size scan), with charge-sensor response data stored as $(30\times30)$ pixel maps from the space of plunger gates (for examples of simulated single- and double-dot regions, respectively, see the right-hand column in Fig.~\ref{fig:dots-examples}). The labels for each measurement are assigned based on the probability of each state within a given realization, i.e., based on the fraction of pixels in each of the three possible states:
\begin{equation}\label{eq:prob_vec}
\begin{split}
\text{\bf p}(V_\mathcal{R})&=[p_{\text{none}},\,p_{\text{SD}},\,p_{\text{DD}}] \\
&=\left[\frac{N-(|\text{SD}|+|\text{DD}|)}{N},\frac{|\text{SD}|}{N},\frac{|\text{DD}|}{N}\right]
\end{split}
\end{equation}
where $|\text{SD}|$ and $|\text{DD}|$ are the numbers of pixels with a single-dot and a double-dot state label, respectively, and $N$ is the size of the image $V_\mathcal{R}$ in pixels. As such, $\text{\bf p}(V_\mathcal{R})$ can be thought of as a probability vector that a given measurement captures each of the possible states (i.e., no dot, single dot, or double dot). The resulting probability vector for a given region $V_\mathcal{R}$, $\text{\bf p}(V_\mathcal{R})$, is an implicit function of the plunger gate voltages defining $V_\mathcal{R}$. It is important to note that, while CNNs are traditionally used to simply classify images into a number of predefined global classes (which can be thought of as a {\it qualitative classification}), we use the raw probability vectors returned by the CNN (i.e., {\it quantitative classification}). 

The CNN architecture consists of two convolutional layers (each followed by a pooling layer) and four fully connected layers with 1024, 512, 256, and 3 units, respectively. The convolutional and pooling layers are used to reduce the size of the feature maps while extracting the most important characteristics of the data. The fully connected layers, on the other hand, allow for nonlinear combinations of these characteristics and classification of the data. We use the Adam optimizer~\cite{Diederik14-AO} with a learning rate $\eta=0.001$, $5000$ steps per training, and a batch size of $50$. The accuracy of the network on the test set is $97.7\,\%$.

%%%%%%%%%%%%%%%%%%%%%%%%%%%%%%%%%%%%%%%%%%%%%%%%%%%%%%%%%%%%%%
\subsection{Optimization and autotuning}\label{subsec:autotuning}
%%%%%%%%%%%%%%%%%%%%%%%%%%%%%%%%%%%%%%%%%%%%%
The optimization step of the autotuning process (Step 4 in Fig.~\ref{fig:tuning}) involves minimization of a fitness function that quantifies how close a probability vector returned by the CNN, $\text{\bf p}(V_\mathcal{R})$, is to the desired vector, $\text{\bf p}_{\text{target}}$. We use a modified version of the original fitness function proposed in Ref.~\cite{Kalantre17-MLD} to include a penalty for tuning to single-dot and no-dot regions:
\begin{equation}\label{eq:fit-function}
\delta(\text{\bf p}_{\text{target}},\bm{p}(V_\mathcal{R})) = \|\text{\bf p}_{\text{target}}-\text{\bf p}(V_\mathcal{R})\|_2 + \gamma(V_\mathcal{R}),
\end{equation}
where $\|\cdot\|_2$ is the $L^2$ norm and the penalty function $\gamma$ is defined as
\begin{equation}
\gamma(V_\mathcal{R}) = \alpha g(p_{\text{none}}) + \beta g(p_{\text{SD}}),
\end{equation}
where $g(x)$ is the arctangent shifted and scaled to assure that the penalty is non-negative [i.e., $g(x) \ge0$] and  that the increase in penalty is more significant once a region is classified as predominantly non-double dot (i.e., the inflection point is at $x=0.5$). Parameters $\alpha$ and $\beta$ are used to weight penalties coming from no dot and single dot, respectively. 

For optimization, we use the Nelder–Mead method~\cite{Nelder65-NMA,Gao12-IMN} implemented in Python~\cite{scipy}. The Nelder-Mead algorithm works to find a minimum of an objective function by evaluating it at initial simplex points---a triangle in the case of the 2D gate space in this work. Depending on the values of the objective function at the simplex points, the subsequent points are selected to move the overall simplex toward the function minimum. In our case, the initial simplex is defined by the fitness value of the starting region $V_\mathcal{R}$ and two additional regions obtained by lowering the voltage on each of the plungers one at a time by $75\,$\si{\milli\volt}.

%%%%%%%%%%%%%%%%%%%%%%%%%%%%%%%%%%%%%%%%%%%%%%%%%%%%%%%%%%%%%
\section{Autotuning the device {\it in situ}}\label{sec:tuning-insitu}
%%%%%%%%%%%%%%%%%%%%%%%%%%%%%%%%%%%%%%%%%%%%%%%%%%%%%%%%%%%%%
To evaluate the autotuner in an experimental setup, a Si/Si$_{x}$Ge$_{1-x}$ quadruple quantum-dot device (see Fig.~\ref{fig:tuning}, Step 1) is precalibrated into an operational mode, with one double quantum dot and one sensing dot active. The evaluation is carried out in  there main phases. In the first phase, we develop a communication protocol between the autotuning software~\cite{qf-lite} and the software used to control the experimental apparatus~\cite{labber}. In the process, we collect 83 measurement scans that are then used to refine the filtering protocol used in Step 2 (see the middle column in Fig.~\ref{fig:dots-examples}). These scans are also used to test the classification accuracy for the neural network. 

\begin{figure}[t]
\includegraphics[width=0.99\linewidth]{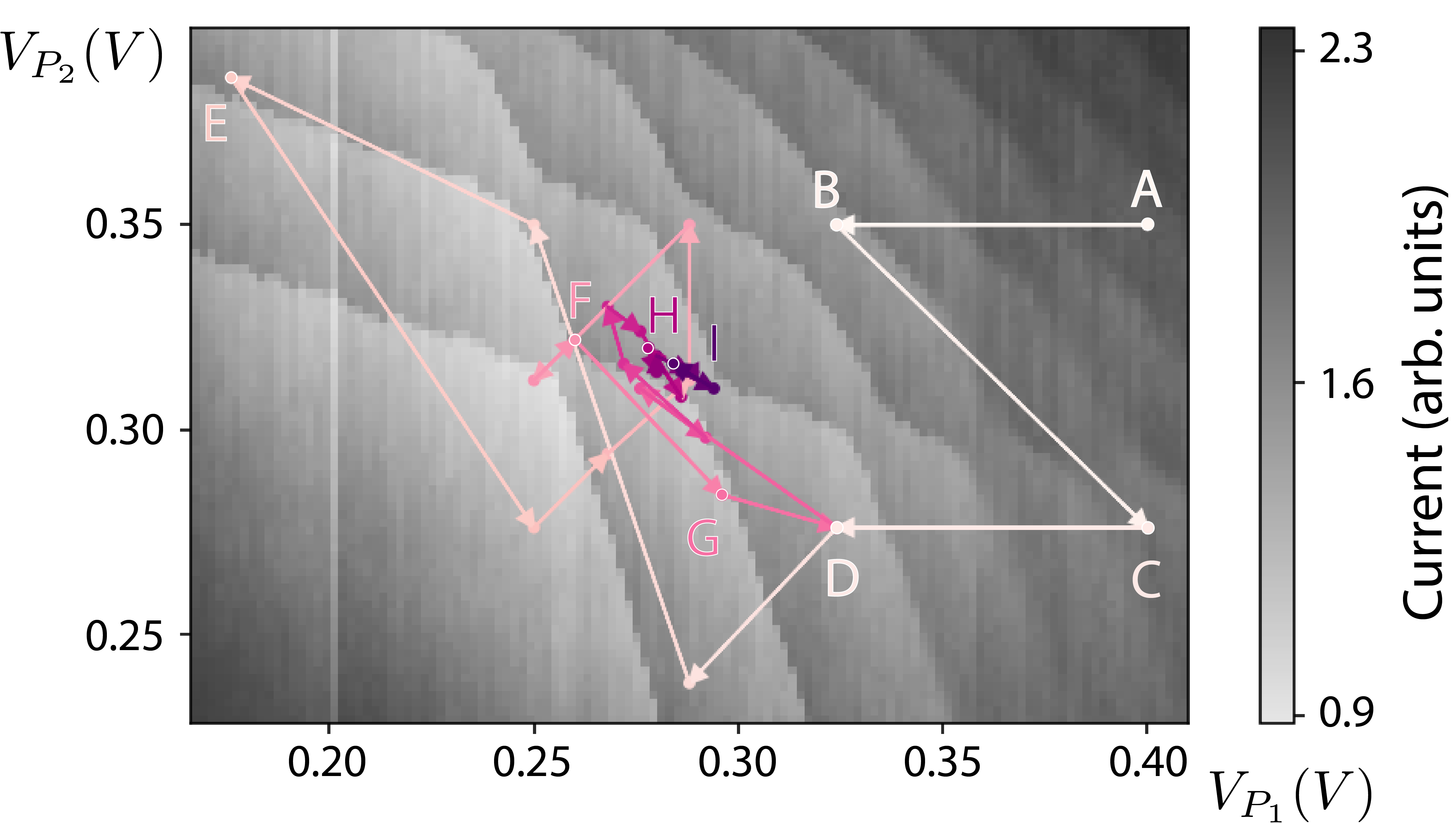}
\caption{An overview of a sample run of the autotuning protocol in the space of plunger gates $(V_{P_1},V_{P_2})$. The arrows and the intensity of the color indicate the progress of the autotuner. The palette corresponds to colors used in the fitness-function plot in Fig.~\ref{fig:sample_run}.}
\label{fig:tuning_path}
\end{figure}

In the second phase, we evaluate the performance of the trained network on hand-labeled experimental data. The data set includes $(30\times 30)\,$\si{\milli \volt} scans with $1\,$\si{\milli \volt} per pixel and $(60\times60)\,$\si{\milli \volt} with $2\,$\si{\milli \volt} per pixel. Prior to analysis, all scans are flattened with an automated filtering function to assure compatibility with the neural network (see the left-hand column in Fig.~\ref{fig:dots-examples}). The accuracy of the trained network in distinguishing between single-dot, double-dot, and no-dot patterns is $81.9\,\%$. 

In the third phase, we perform a series of trial runs of the autotuning algorithm in the $(V_{P_1}, V_{P_2})$ plunger space, as shown in Fig.~\ref{fig:sample_run}. To prevent tuning to voltages outside of the device tolerance regime, we sandbox the tuner by limiting the allowed plunger values to between $0$ and $600\,$\si{\milli\volt}. Attempts to perform measurements outside of these boundaries during a tuning run are blocked and a fixed value of $2$ (i.e., a maximum fit value) is assigned to the fitness function.

We initialize 45 autotuning runs, out of which seven are terminated by the user due to technical problems (e.g., stability of the sensor). Of the remaining 38 completed runs, in 13 cases the scans collected at an early stage of the tuning process are found to be incompatible with the CNN. In particular, while there are three possible realizations of the single-dot state (coupled strongly to the left plunger, the right plunger, or equally coupled, forming a ``central dot''), the training data set includes predominantly realizations of the ``central dot'' state. As a result, whenever the single left or right plunger dot is measured, the scan is labeled incorrectly. When a sequence of consecutive ``single-plunger-dot'' scans is used in the optimization step, the optimizer misidentifies the scans as double dot and fails to tune away from this region. These runs are removed from further analysis, as with the incorrect labels, the autotuner terminates each time in a region classified as double dot (i.e., a success from the ML perspective) which in reality is a single dot (i.e., a failure for practical purposes). We discuss the performance of the autotuner based on the remaining 25 runs. 

\begin{table}[tp]
\renewcommand{\arraystretch}{1.1}
\renewcommand{\tabcolsep}{4pt}
\caption{A summary of the performance for the experimental test runs ($N_{\text{tot}} = 14$). $N_{\text{exp}}$ denotes the number of experimental runs initiated at position $(V_{P_1},V_{P_2})$ (\si{\milli\volt}), $N_{\text{suc}}$ indicates the number of successful experimental runs, and $P_{\Delta=75}$ (\%), $P_{\Delta=100}$ (\%), and $P_{\Delta=f(\delta_0)}$ (\%) are the success rates for the 81 test runs with optimization parameters resembling the experimental configuration (fixed simplex size $\Delta=75$~\si{\milli\volt}), with the initial simplex size increased to $100$~\si{\milli\volt}, and with initial simplex size dynamically adjusted based on the fitness value of the first scan, respectively. All test runs are performed using the new neural network.} 
\centering
\begin{tabular}{cccrrr}
\hline \hline
$(V_{P_1},V_{P_2})$ &  $N_{\text{exp}}$ & $N_{\text{suc}}$ & $P_{\Delta=75}$ & $P_{\Delta=100}$ & $P_{\Delta=f(\delta_0)}$\\
\hline
(250,400) & 1 & 1 & 85.2 & 100.0 & 93.8 \\
(350,400) & 6 & 6 & 74.1 &  95.1 & 95.1 \\
(350,415) & 1 & 0 & 75.3 &  86.4 & 96.3 \\
(350,425) & 1 & 1 & 55.6 &  86.4 & 85.2 \\
(350,450) & 3 & 2 &  3.7 &  18.5 & 34.6 \\
(400,350) & 1 & 1 &  4.9 &  69.1 & 93.8 \\
(450,350) & 1 & 1 & 17.3 &   1.2 & 23.5 \\
\hline \hline
\label{tab:performance}
\end{tabular}
\end{table}

While tuning, it is observed that the autotuner tends to fail when initiated further away from the target double-dot region. An inspection of the test runs confirms that whenever both plungers are set at or above $375\,$\si{\milli \volt}, the tuner becomes stuck in the  plateau area of the fitness function and does not reach the target area (with two exceptions). Out of the 25 completed runs, 14 are initiated with at least one plunger set below $375\,$\si{\milli \volt}. Out of these, two cases fail, both due to instability of the charge sensor resulting in unusually noisy data that is incorrectly labeled by the CNN and thus leads to an inconsistent gradient direction. The overall success rate here is $85.7\,\%$ (for a summary of the performance for each initial point from this class, see Table~\ref{tab:performance}). When both plungers are set at or above $375\,$\si{\milli \volt}, only 2 out of 11 runs are successful ($18.2\,\%$), with all failing cases resulting from ``flatness'' of the fit function [for a visualization of the fitness function over a large range of voltages in the space of plunger gates $(V_{P_1},V_{P_2})$, see Fig.~\ref{fig:fit_fun_vis}].

%%%%%%%%%%%%%%%%%%%%%%%%%%%%%%%%%%%%%%%%%%%%%%%%%%%%%%%%%%%%%
\section{``Off-line'' tuning}\label{sec:tuning-offline}
%%%%%%%%%%%%%%%%%%%%%%%%%%%%%%%%%%%%%%%%%%%%%%%%%%%%%%%%%%%%%
Tuning ``off-line''---tuning within a premeasured scan for a large range of gate voltages that captures all possible state configurations---allows for the study of how the various parameters of the optimizer impact the functioning of the autotuner and the further investigation of the reliability of the tuning process while not taking up experimental time. The scan that we use for off-line tuning spans 125~\si{\milli\volt} to 525~\si{\milli\volt} for plunger $P_1$ and 150~\si{\milli\volt} to 550~\si{\milli\volt} for $P_2$, measured in $2\,$\si{\milli\volt} per pixel resolution. 

The deterministic nature of the CNN classification (i.e., assigning a fixed probability to a given scan) assures that the performance of the tuner will be affected solely by changes made to the optimizer. On the other hand, with static data, for any starting point the initial simplex and the consecutive steps are fully deterministic, making a reliability test challenging. To address this issue, rather than repeating a number of autotuning tests for a given starting point $(V_{P_1},V_{P_2})$, we initiate tuning runs for points sampled from a $(9\times9)$ pixels region around $(V_{P_1},V_{P_2})$, resulting in 81 test runs for each point.

\begin{figure}[pt]
\includegraphics[width=0.99\linewidth]{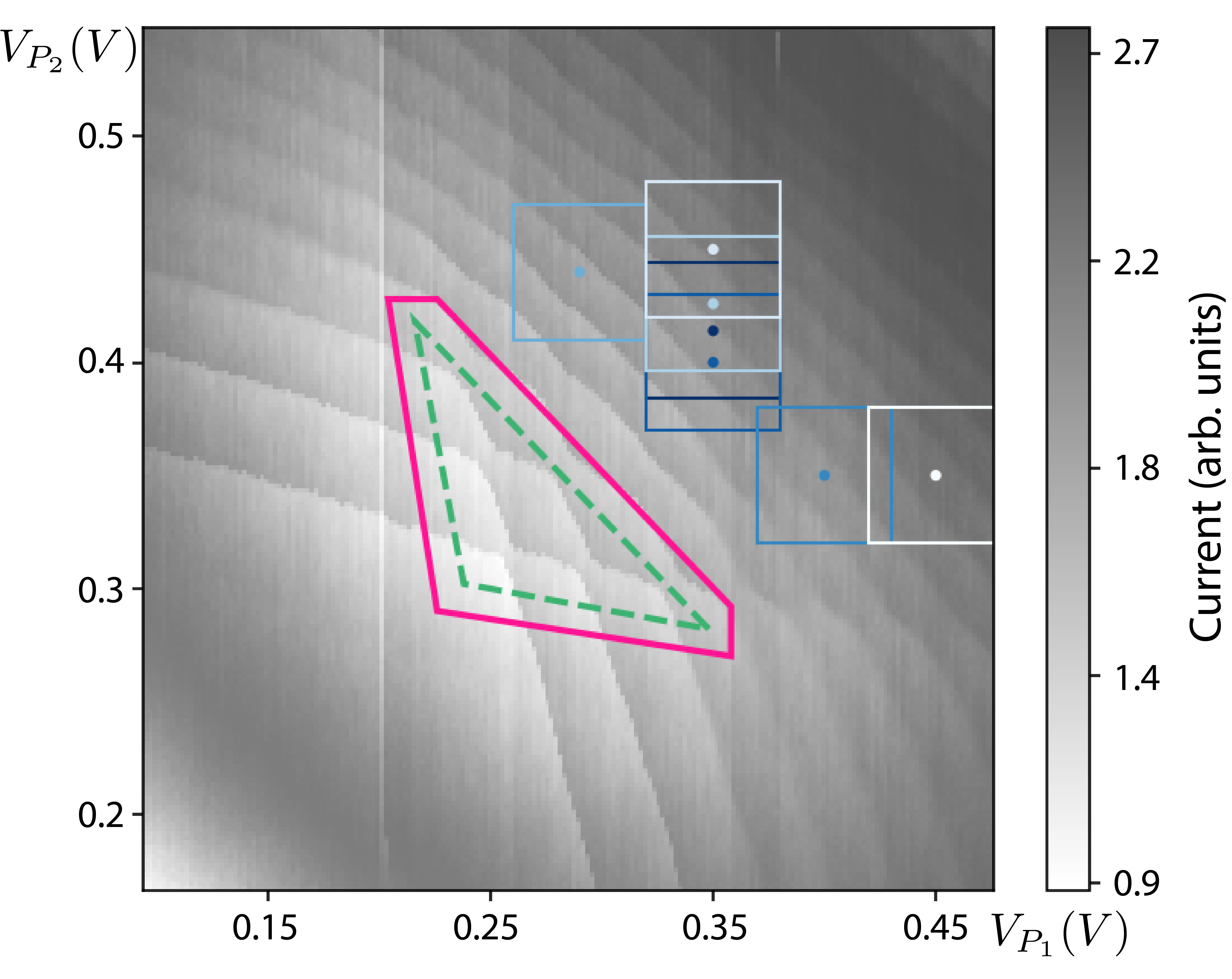}
\caption{Visualizations of the ``ideal'' (marked with dashed green triangle) and the ``sufficiently close'' (marked with solid magenta diamond) regions used to determine the success rate for the off-line tuning. All considered initial regions listed in Table~\ref{tab:performance} are marked with squares. The intensities of the colors correspond to the success rate when using dynamic simplex (a darker color denotes a higher success rate).}
\label{fig:exp_vis}
\end{figure}

We assess the reliability of the autotuning protocol for the seven experimentally tested configurations listed in Table~\ref{tab:performance} [note that for point $(250,400)\,$\si{\milli \volt}, the gate values are adjusted when testing over the premeasured scan to account for changes in the screening gates]. To quantify the performance of the tuner, we define the tuning success rate, $P$, as a fraction of runs that ended in the ``ideal'' region (marked with a green triangle in Fig.~\ref{fig:exp_vis}) or in the ``sufficiently close'' region (marked with a magenta diamond in Fig.~\ref{fig:exp_vis}) with weights $1$ and $0.5$, respectively. Moreover, in the network analysis step, we use a neural network with the same architecture as discussed in Sec.~\ref{subsec:ml} but trained on a new data set that includes all three realizations of the SD state. When using optimization parameters resembling those implemented in the laboratory (i.e., fixed simplex of a size $\Delta=75\,$\si{\milli\volt}) and a new neural network, the overall success rate is $45.2\,\%$ with standard a deviation (s.d.) of $35.5\,\%$. The summary of the performance for each point is presented in Table~\ref{tab:performance} (for a comparison of the number of iterations between points, see Table~\ref{tab:tuning_parameters}). Increasing the initial simplex size by $25\,$\si{\milli\volt} significantly improves the success rate for all but two points (see the $P_{\Delta=100}$ column in Table~\ref{tab:performance}), with the overall success rate of $65.2\,\%$ ($\text{s.d.} = 39.4\,\%$). The $P_{\Delta=f(\delta_0)}$ column in Table~\ref{tab:performance} shows the success rate for tuning when the initial simplex size is scaled based on the fitness value of the initial step $\delta_0$, such that tuning from points further away from the target area will use a larger simplex than those initiated relatively close to the ``ideal'' region. The overall success rate here is $74.6\,\%$ ($\text{s.d.} = 31.5\,\%$).

\begin{figure}[pt]
\includegraphics[width=0.99\linewidth]{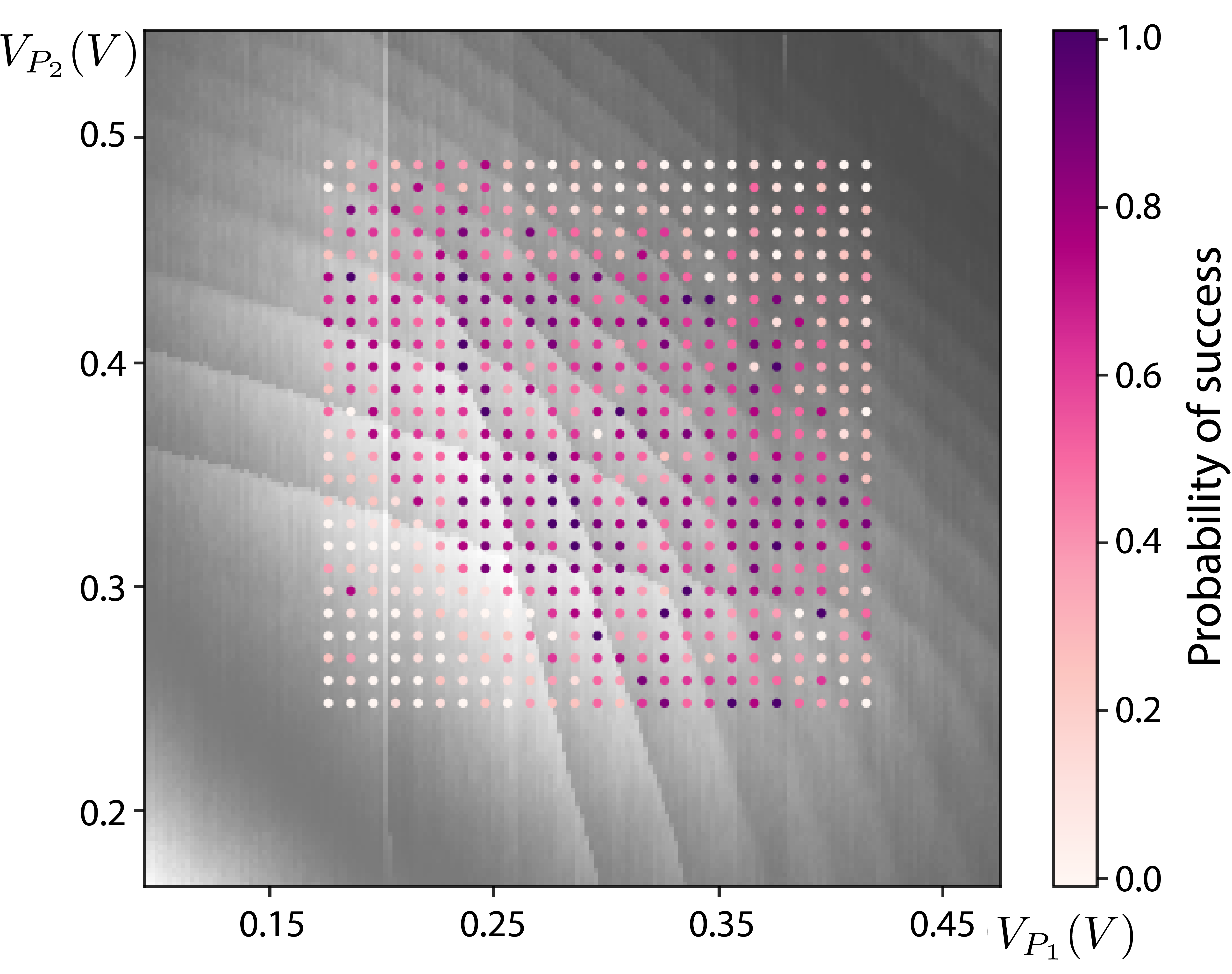}
\caption{A heat map of the probability of success when tuning off-line over a set of $N=4$ premeasured devices. The intensity of the colors corresponds to the success rate, with a darker color denoting a higher success rate.}
\label{fig:heat-map}
\end{figure}

Finally, to assess the performance of the autotuning protocol for a wider range of initial configurations, we perform off-line tuning over a set of premeasured scans. Using four scans spanning 100~\si{\milli\volt} to 500~\si{\milli\volt} for plunger $P_1$ and 150~\si{\milli\volt} to 550~\si{\milli\volt} for $P_2$, measured in $2\,$\si{\milli\volt} per pixel resolution, we initiate $N=784$ test runs per scan, sampling every $10\,$\si{\milli\volt} and leaving a margin that is big enough to ensure that the initial simplex is within the full scan boundaries. A heat map representing the performance of the autotuner is presented in Fig.~\ref{fig:heat-map}. As can be seen, the autotuner is most likely to fail when initiated with both plunger gates set to either high (above $400\,$\si{\milli\volt}) or low (below $300\,$\si{\milli\volt}) voltage. While in both cases the ``flatness'' of the fitness function contributes to the tuning failure, the fixed direction of the initial simplex further contributes to this issue. Adding rotation to the simplex, i.e., varying both plunger gates when determining the second and third steps in the optimization (see B and C in Fig.~\ref{fig:tuning_path}), may help with the latter problem.

%%%%%%%%%%%%%%%%%%%%%%%%%%%%%%%%%%%%%%%%%%%%%%%%%%%%%%%%%%%%%
\section{Summary and Outlook}\label{sec:summary}
%%%%%%%%%%%%%%%%%%%%%%%%%%%%%%%%%%%%%%%%%%%%%%%%%%%%%%%%%%%%%
While a standardized fully automated approach to tuning quantum-dot devices is essential for their scalability, present-day approaches to tuning rely heavily on human heuristic and algorithmic protocols that are specific to a particular device and cannot be used across devices without fine readjustments. To address this issue, we are developing a tuning paradigm that combines synthetic data from a physical model with ML and optimization techniques to establish an automated closed-loop system of experimental device control. Here, we report on the performance of the proposed autotuner when tested {\it in situ}.

In particular, we verify that, within certain constraints, the proposed approach can automatically tune a QD device to a desired double-dot configuration. In the process, we confirm that a ML algorithm, trained using exclusively synthetic noiseless data, can be used to successfully classify images coming from experiment, where noise and imperfections typical of real measurements are present.This work also enables us to identify areas in which further work is necessary to improve the overall reliability of the autotuning system. A new training data set is necessary to account for all three possible single-dot states. The size of the initial simplex also seems to contribute to the mobility of the tuner out of the SD plateau. For comparison, in Table~\ref{tab:performance} we present the performance of a tuner using the new network and a bigger simplex size for the experimentally tested starting points. In terms of the length of the tuning runs, at present, the bottleneck of the protocol is the time it takes to perform scans (about 5 min per scan) and the repeated iterations toward the termination of the cycle (i.e., repeated scans of the same region). This can be improved by orders of magnitude by using faster voltage sources and readout techniques and by developing a custom optimization algorithm. Regardless, the power of this technique lies in its automation, allowing a skilled researcher to spend time elsewhere.

These results serve as a baseline for future investigation of fine-grain device control (i.e., tuning to desired charge configuration) and of ``cold-start'' autotuning (i.e., complete tuning without any precalibration of the device). Finally, our work paves the way for similar approaches applied to a wide range of experiments in physics.

To use QD qubits in quantum computers, it is necessary to develop a reliable automated approach to control QD devices, independent of human heuristics and intervention. Working with experimental devices with high-dimensional parameter spaces poses many challenges, from performing reliable measurements to identifying the device state to tuning into a desirable configuration. By combining theoretical, computational, and experimental efforts, this interdisciplinary research sheds light on how modern ML techniques can assist experiments.

%%%%%%%%%%%%%%%%%%%%%%%%%%%%%%%%%%%%%%%%%%%%%%%%%%%%%%%%%%%%%%
\section*{Acknowledgements}\label{sec:acknowledge}
%%%%%%%%%%%%%%%%%%%%%%%%%%%%%%%%%%%%%%%%%%%%%%%%%%%%%%%%%%%%%%
This research was sponsored in part by the Army Research Office (ARO), through Grant No. W911NF-17- 1-0274. The development and maintenance of the growth facilities used for fabricating samples was supported by the Department of Energy, through Grant No. DE-FG02- 03ER46028. We acknowledge the use of facilities sup- ported by the National Science Foundation through the University of Wisconsin-Madison Materials Research Science and Engineering Center (Grant No. DMR-1720415). S.K. gratefully acknowledges support from the Joint Quantum Institute –Joint Center for Quantum Information and Computer Science Lanczos graduate fellowship. The views and conclusions contained in this paper are those of the authors and should not be interpreted as rep- resenting the official policies, either expressed or implied, of the ARO, or the U.S. Government. The U.S. Government is authorized to reproduce and distribute reprints for Government purposes notwithstanding any copyright notation herein. Any mention of commercial products is for information only; it does not imply recommendation or endorsement by the National Institute of Standards and Technology.

%%%%%%%%%%%%%%%%%%%%%%%%%%%%%%%%%%%%%%%%%%%%%%%%%%%%%%%%%%%%%%
\appendix
%\renewcommand\thefigure{\thesection.\arabic{figure}}    
%\renewcommand\thetable{\thesection.\Roman{table}}    
%%%%%%%%%%%%%%%%%%%%%%%%%%%%%%%%%%%%%%%%%%%%%%%%%%%%%%%%%%%%%
\section{Data processing}
%%%%%%%%%%%%%%%%%%%%%%%%%%%%%%%%%%%%%%%%%%%%%%%%%%%%%%%%%%%%%
\begin{figure}[b]
\includegraphics[width=0.99\linewidth]{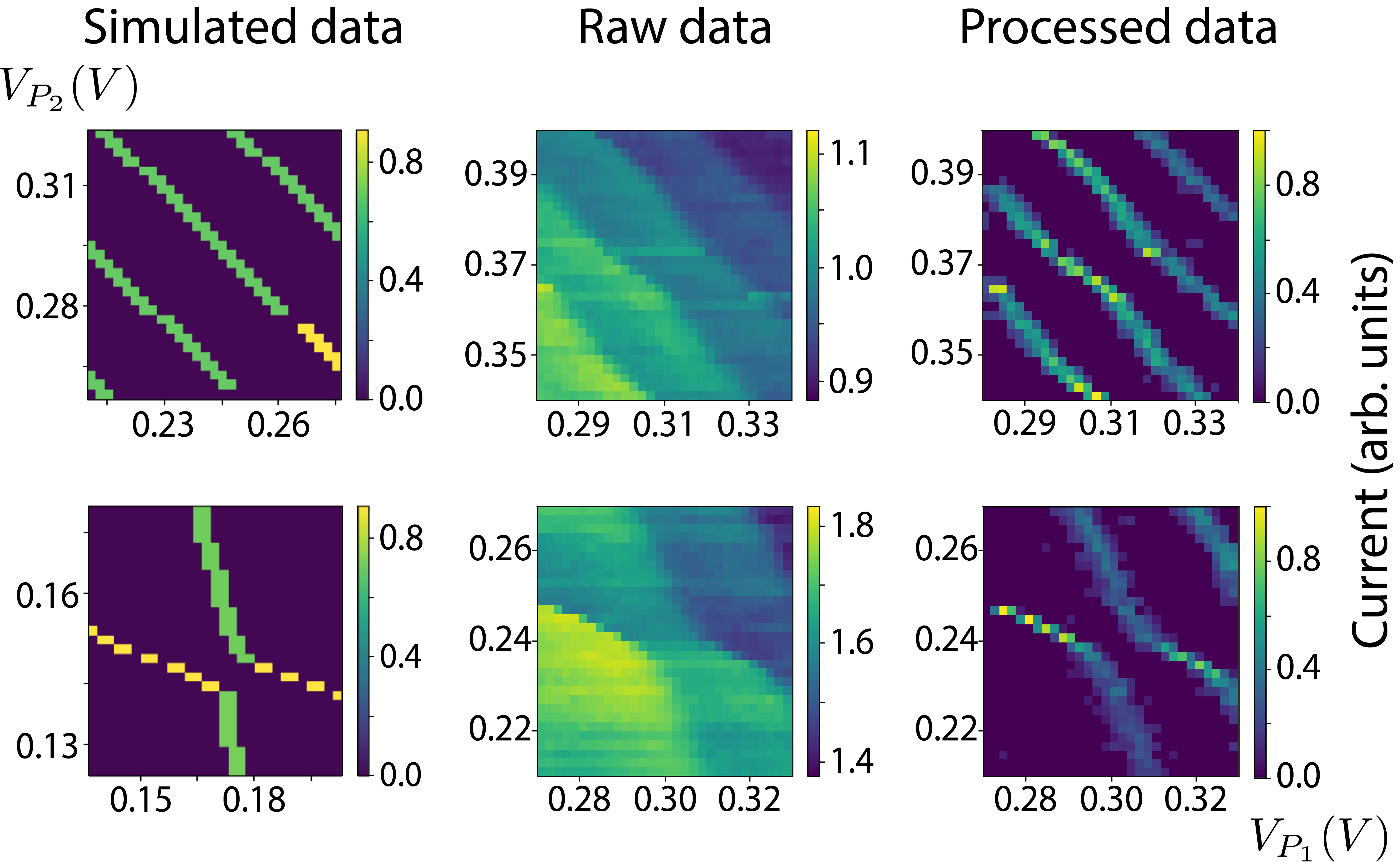}
\caption{The relationship between the simulated, raw, and processed data. The top row consists of sample scans with single-dot regions and the bottom row of scans with double-dot regions. The left-hand column shows the simulated data, the middle column shows the raw acquired experimental data, and the right-hand column shows the processed experimental data (as ``seen'' by the CNN classifier).\label{fig:dots-examples}}
\end{figure}

The model used to simulate the QD devices~\cite{Zwolak18-QLD} does not account for the noise present in a real measurement. As a result, the data used to train the CNN classifier, obtained by taking a numerical gradient of the sensor data, lead to very clean data, with the background uniformly flattened and charge-transition lines clearly visible (see the first column in Fig.~\ref{fig:dots-examples}). To assure compatibility with the CNN classifier, the acquired experimental scans need to be processed before the probability vector can be assigned to it. Here, the data processing consists of three steps: the numerical derivative, followed by thresholding and resizing. To minimize noise, the derivative is taken in the direction of measurement. The gradient data is also tested against unexpected charge-sensor flipping and, if necessary, reverted to assure positive values at the charge-transition lines. An automated protocol is implemented to normalize the data and to remove the background noise. Finally, the data is resized to $(30\times30)$ pixels resolution. The second and third columns in Fig.~\ref{fig:dots-examples} show sample raw and processed data, respectively, for a single- and double-dot image.

%%%%%%%%%%%%%%%%%%%%%%%%%%%%%%%%%%%%%%%%%%%%%%%%%%%%%%%%%%%%%
\section{Effect of simplex size on off-line tuning}
%%%%%%%%%%%%%%%%%%%%%%%%%%%%%%%%%%%%%%%%%%%%%%%%%%%%%%%%%%%%%
\begin{table}[t]
\renewcommand{\arraystretch}{1.1}
\renewcommand{\tabcolsep}{6pt}
\caption{The average (standard deviation in parentheses) number of iterations when tuning off-line for varying configurations of the initial simplex $\Delta$. In all cases, the average is taken over $N=81$ test runs for points sampled within $10\,$\si{\milli\volt} around each experimentally tested point given by $(V_{P_1},V_{P_2})$. } 
\centering
\begin{tabular}{crrr}
\hline \hline
$(V_{P_1},V_{P_2})$ &  $\Delta=75\,$\si{\milli\volt} & $\Delta=100\,$\si{\milli\volt} & $\Delta=f(\delta_0)$ \\
\hline
(250,400) & 12.7 (2.5) & 12.2 (2.0) & 12.6 (2.2) \\
(350,400) & 14.0 (2.4) & 13.6 (2.2) & 13.5 (2.3) \\
(350,415) & 13.2 (2.3) & 14.1 (2.1) & 13.4 (2.1) \\
(350,425) & 12.9 (2.3) & 13.9 (2.1) & 13.6 (2.2) \\
(350,450) & 11.6 (2.7) & 13.3 (2.4) & 13.9 (2.5) \\
(400,350) & 13.9 (2.3) & 14.0 (2.2) & 13.3 (1.8) \\
(450,350) & 14.5 (2.6) & 15.0 (2.6) & 15.0 (2.5) \\
\hline \hline
\label{tab:tuning_parameters}
\end{tabular}
\end{table}

While varying the simplex size significantly affects the performance of the autotuner, leading to an increase in the overall accuracy for the tested points by nearly $40\,\%$ (for details, see Table~\ref{tab:performance}), it does not affect the number of iterations of the optimizer. In particular, the overall average numbers of iterations for the three tested simplex sizes are as follows: $13.3$ (pooled $\text{s.d.} = 2.5$), $13.7$ (pooled $\text{s.d.} = 2.3$), and $13.6$ (pooled $\text{s.d.} = 2.3$) for tuning with an initial size of $\Delta=75\,$\si{\milli\volt}, $\Delta=100\,$\si{\milli\volt}, and $\Delta=f(\delta_0)$, respectively. Table~\ref{tab:tuning_parameters} shows the average numbers of iterations executed by the optimizer for each tested point.

%%%%%%%%%%%%%%%%%%%%%%%%%%%%%%%%%%%%%%%%%%%%%%%%%%%%%%%%%%%%%
\section{Fitness function}
\begin{figure}[b]
\includegraphics[width=0.99\linewidth]{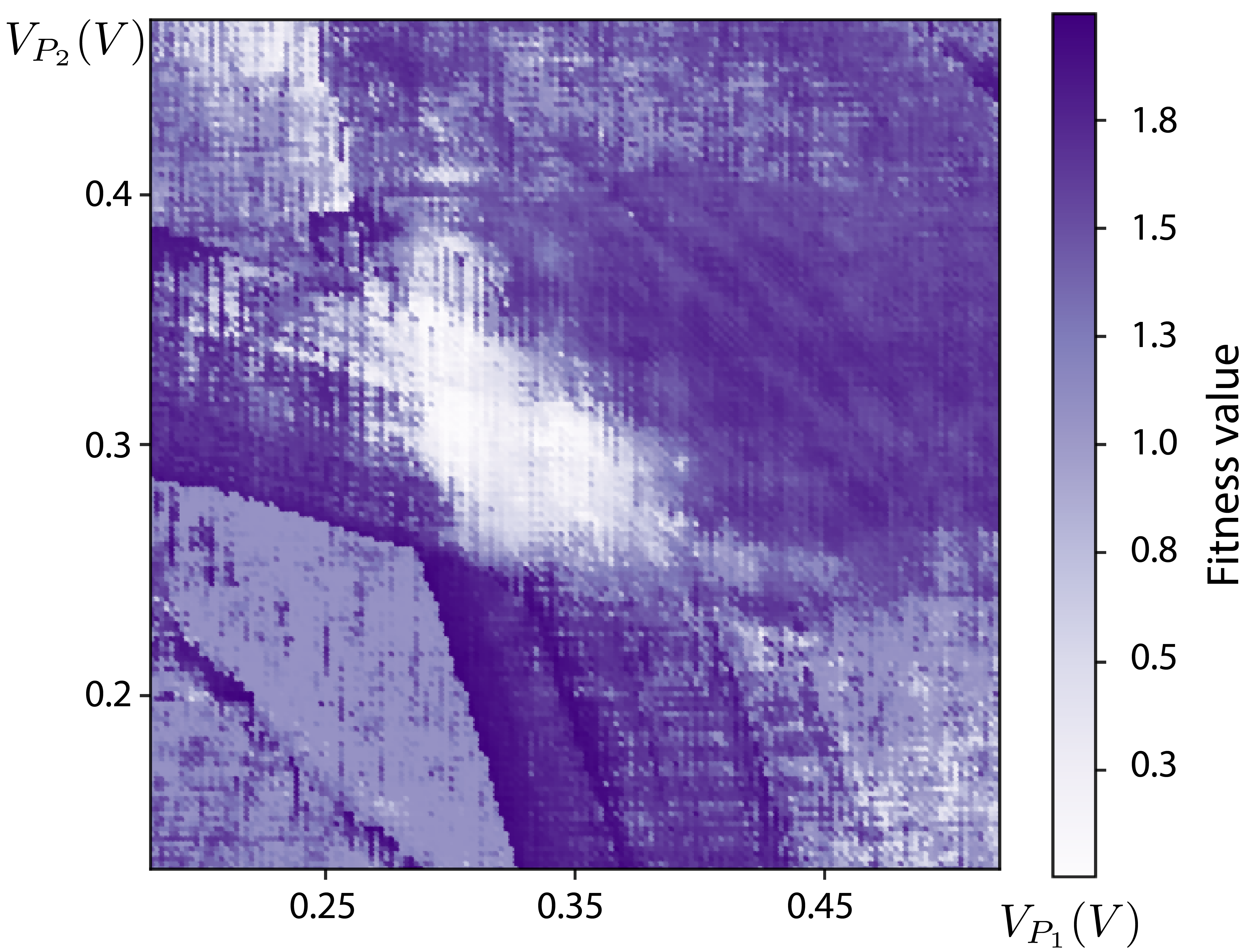}
\caption{The fitness function over a sample device shown in Fig.~\ref{fig:exp_vis}.  \label{fig:fit_fun_vis}}
\end{figure}

%%%%%%%%%%%%%%%%%%%%%%%%%%%%%%%%%%%%%%%%%%%%%%%%%%%%%%%%%%%%%
%\begin{minipage}[b]{1.0\linewidth}
We plot the fitness value for tuning to a double-dot regime as a function of the plunger gate voltages for a scan with experimental data. In particular, for each point in the voltage space, as presented in Fig.~\ref{fig:exp_vis}, we calculate the fitness value for a region centered at this point. This allows us to represents the landscape over which the autotuning optimization runs [a $(171 \times 171)$ pixels map]. One can see the double-dot state forming a minimum near the center of Fig.~\ref{fig:fit_fun_vis}, which represents the target area for tuning. 
%\end{minipage}

%%%%%%%%%%%%%%%%%%%%%%%%%%%%%%%%%%%%%%%%%%%%%%%%%%%%%%%%%%%%%%

%%%%%%%%%%%%%%%%%%%%%%%%%%%%%%%%%%%%%%%%%%%%%%%%%%%%%%%%%%%%%%
\end{document}